\documentclass[12pt]{article}
\usepackage{graphicx}
\usepackage{amsmath,amsfonts,amssymb}
\usepackage{hyperref}

\newcommand\pubnumber{CIPANP2018-DeTar}
\newcommand\pubdate{\today}

\newcommand*{\MeV}{{\rm Me\!V}}
\newcommand*{\GeV}{{\rm Ge\!V}}

\newcommand*{\chpt}{\raise0.4ex\hbox{$\chi$}PT}
\newcommand*{\schpt}{S\raise0.4ex\hbox{$\chi$}PT}

\newcommand*{\etc}{\textit{etc.}}

\newcommand{\fpiPDG}{f_{\pi,{\rm PDG}}}
\newcommand{\vt}{\text{x}}

\newcommand{\MRSt}{\text{MRS}}

\newcommand{\MSbar}{{\overline{\text{MS}}} }

\newcommand{\mbar}{{\overline{m}} }

\newcommand{\be}{\begin{equation}}
\newcommand{\ee}{\end{equation}}
\newcommand{\bea}{\begin{eqnarray}}
\newcommand{\eea}{\end{eqnarray}}

\newcommand{\cO}{\ensuremath{\mathcal{O}}}

\def\Title#1{\begin{center} {\Large #1 } \end{center}}
\def\Author#1{\begin{center}{ \sc #1} \end{center}}
\def\Address#1{\begin{center}{ \it #1} \end{center}}

\newcommand\pubblock{\rightline{\begin{tabular}{l} \pubnumber\\
         \pubdate  \end{tabular}}}
\newenvironment{Abstract}{\begin{quotation}  }{\end{quotation}}
\newenvironment{Presented}{\begin{quotation} \begin{center} 
             PRESENTED AT\end{center}\bigskip 
      \begin{center}\begin{large}}{\end{large}\end{center} \end{quotation}}
\def\Acknowledgements{\bigskip  \bigskip \begin{center} \begin{large}
             \bf ACKNOWLEDGEMENTS \end{large}\end{center}}

\def\beq{\begin{equation}}
\def\eeq#1{\label{#1}\end{equation}}
\def\eeqn{\end{equation}}

\def\beqa{\begin{eqnarray}}
\def\eeqa#1{\label{#1}\end{eqnarray}}
\def\eeqan{\end{eqnarray}}

\let\bar=\overbar

\def\Dslash{\not{\hbox{\kern-4pt $D$}}}
\def\dslash{\not{\hbox{\kern-2pt $\del$}}}

\def\ee{e^+e^-}

\def\msb{{\bar{\ssstyle M \kern -1pt S}}}

\begin{document}
\begin{titlepage}
\pubblock

\vfill
\Title{B- and D-meson leptonic decay constants and quark masses from four-flavor lattice QCD}
\vfill
\Author{A.~Bazavov$^a$, C.~Bernard$^b$, N.~Brambilla$^{cd}$, N.~Brown$^b$, C. DeTar$^e$\footnote{Speaker}, }
\Author{A.X. El Khadra$^{fg}$, E.~G\'amiz$^h$, Steven Gottlieb$^i$, U.M.~Heller$^j$,}
\Author{J.~Komijani$^{cdk}$, A.S.~Kronfeld$^{gd}$,J.~Laiho$^l$, P.M.~Mackenzie$^g$, E.T.~Neil$^{mn}$,}
\Author{ J.N.~Simone$^g$, R.L.~Sugar$^o$, D.~Toussaint$^p$, R.S. Van de Water$^g$, A. Vairo$^c$,}
\Address{$^a$Department of Computational Mathematics, Science and Engineering,
and Department of Physics and Astronomy, Michigan State University, East Lansing, MI 48824, USA}
\Address{$^b$Department of Physics, Washington University, St. Louis, MO 63130, USA}
\Address{$^c$Physik-Department, Technische Universit\"at M\"unchen, James-Franck-Stra{\ss}e 1, 85748 Garching, Germany}
\Address{$^d$Institute for Advanced Study, Technische Universit\"at M\"unchen, Lichtenbergstra{\ss}e 2a, 85748 Garching, Germany}
\Address{$^e$Department of Physics and Astronomy, University of Utah, Salt Lake City, UT 84112, USA}
\Address{$^f$Department of Physics, University of Illinois, Urbana,  IL 61801, USA}
\Address{$^g$Fermi National Accelerator Laboratory, Batavia, IL 60510, USA}
\Address{$^h$CAFPE and Departamento de F\'isica Te\'orica y del Cosmos, Universidad de Granada, E-18071 Granada, Spain}
\Address{$^i$Department of Physics, Indiana University, Bloomington, IN 47405, USA}
\Address{$^j$American Physical Society, One Research Road, Ridge, NY 11961, USA}
\Address{$^k$School of Physics and Astronomy, University of Glasgow, Glasgow G12 8QQ, United~Kingdom}
\Address{$^l$Department of Physics, Syracuse University, Syracuse, NY 13244, USA}
\Address{$^m$Department of Physics, University of Colorado, Boulder, CO 80309, USA}
\Address{$^n$RIKEN-BNL Research Center, Brookhaven National Laboratory, \\ Upton, New York 11973, USA}
\Address{$^o$Department of Physics, University of California, Santa Barbara, CA 93106, USA}
\Address{$^p$Physics Department, University of Arizona, Tucson, AZ 85721, USA}
\vfill
\begin{Abstract}
We describe a recent lattice-QCD calculation of the leptonic decay
constants of heavy-light pseudoscalar mesons containing charm and
bottom quarks and of the masses of the up, down, strange, charm, and
bottom quarks.  Results for these quantities are of the highest
precision to date.  Calculations use 24 isospin-symmetric
ensembles of gauge-field configurations with six different lattice
spacings as small as approximately 0.03 fm and several values of the
light quark masses down to physical values of the average up- and
down-sea-quark masses.  We use the highly-improved staggered quark
(HISQ) formulation for valence and sea quarks, including the bottom
quark. The analysis employs heavy-quark effective theory (HQET).  A
novel HQET method is used in the determination of the quark masses.
\end{Abstract}
\vfill
\begin{Presented}
Conference on Intersections of Particle and Nuclear Physics\\
Palm Springs, California, 29 May -- 3 June, 2018
\end{Presented}
\vfill
\end{titlepage}
\def\thefootnote{\fnsymbol{footnote}}
\setcounter{footnote}{0}

\section{Introduction}

A promising strategy for discovering new physics compares
high-precision predictions of the standard model with high-precision
experimental measurements, especially for rare higher-order weak
processes.  Precise values of the decay constants $f_B$ and $f_{B_s}$
are needed to provide accurate predictions for rare decays such as
$B\rightarrow \tau\nu$, for neutral $B$ mixing, and for the rare
process $B_s \rightarrow \mu^+\mu^-$.  They are also needed to probe
the $V-A$ structure of the weak $W/u/b$ vertex and to resolve or
sharpen the tension between inclusive and exclusive determinations of
the CKM matrix element $|V_{ub}|$.  Precise values of the quark masses
are needed for precise Standard-Model predictions and to test Higgs
origin of those masses. This talk describes results from two such
high-precision {\it ab initio} studies using lattice quantum
chromodynamics, one devoted to decay constants \cite{Bazavov:2017lyh}
and one, to quark masses \cite{Bazavov:2018omf,Basak:2018yzz}. For
details, please see those references.

Several improvements over previous calculations have allowed us to
achieve high precision.  We use gluon gauge-field configurations
generated with the highly-improved staggered quark (HISQ) formulation
for sea quarks, which reduces significantly light-quark lattice
discretization errors \cite{Follana:2006rc}. We use the same quark
formulation for all valence quarks, including $b$, following the HPQCD
collaboration \cite{McNeile:2011ng}.  We combine three effective field
theories (EFTs) to carry out extrapolations to the physical point,
namely, heavy-quark effective theory (HQET) to treat heavy-quark
discretization effects, heavy-meson rooted, all-staggered chiral
perturbation theory (HMrA\schpt) to treat the light-quark mass
dependence, and Symanzik effective theory (SET) to treat light-quark
and gluon discretization errors.  For the quark-mass calculation, we
use a new minimal-renormalon-subtraction (MRS) scheme to improve the
HQET descriptions of the meson masses in terms of quark masses
\cite{Komijani:2017vep,Brambilla:2017hcq}.  Finally, we have
accumulated a large simulation sample that includes 24 gauge-field
ensembles with approximate lattice spacing ranging from 0.03 to 0.15
fm with four flavors of sea quarks, namely, mass-degenerate up and
down quarks and physical-mass strange and charm quarks. Except for
lattice spacing 0.03 fm, one of the ensembles for each lattice spacing
has all physical-mass sea quarks.  The others have varying ratios of
the mass of the light sea quark and strange sea quark.  For further
details, please see \cite{Bazavov:2017lyh}.

\section{Decay-constant methodology}

Since the study varies the heavy and light valence quark masses, we
use the notation ``$H_\vt$'' instead of ``$B^+$'' or ``$B_s$'' for the
heavy-light pseudoscalar meson. The heavy-light-meson decay constant
is obtained from the amplitude $A_{\rm pt-pt}$ of the point-to-point
pseudoscalar density-density correlator at large Euclidean time $t$:
\begin{equation}
C_{\rm pt-pt}(t) \rightarrow A_{\rm pt-pt} \exp(-M_{H_\vt} t) \,.
\end{equation}
The decay constant $F_{H_\vt}$ is obtained from its square root:
\begin{equation}
F_{H_\vt} = (m_{h} + m_{\vt})\sqrt{\frac{3V A_{pt-pt}}{2M^3_{H_\vt}}} \,.
\end{equation}
This calculation is done with bare light valence-quark masses
$m_{\vt} \in [(m_u + m_d)/2, m_s]$ and bare heavy valence-quark
masses $m_h \in [m_c, m_b]$.  To avoid
large heavy-quark discretization errors, we include results only with
bare heavy-quark masses with approximately $m_h < 0.9/a$ in this study.  

\begin{figure}
  \includegraphics[width=2.5in,trim={0.5cm 0 0 0},clip]{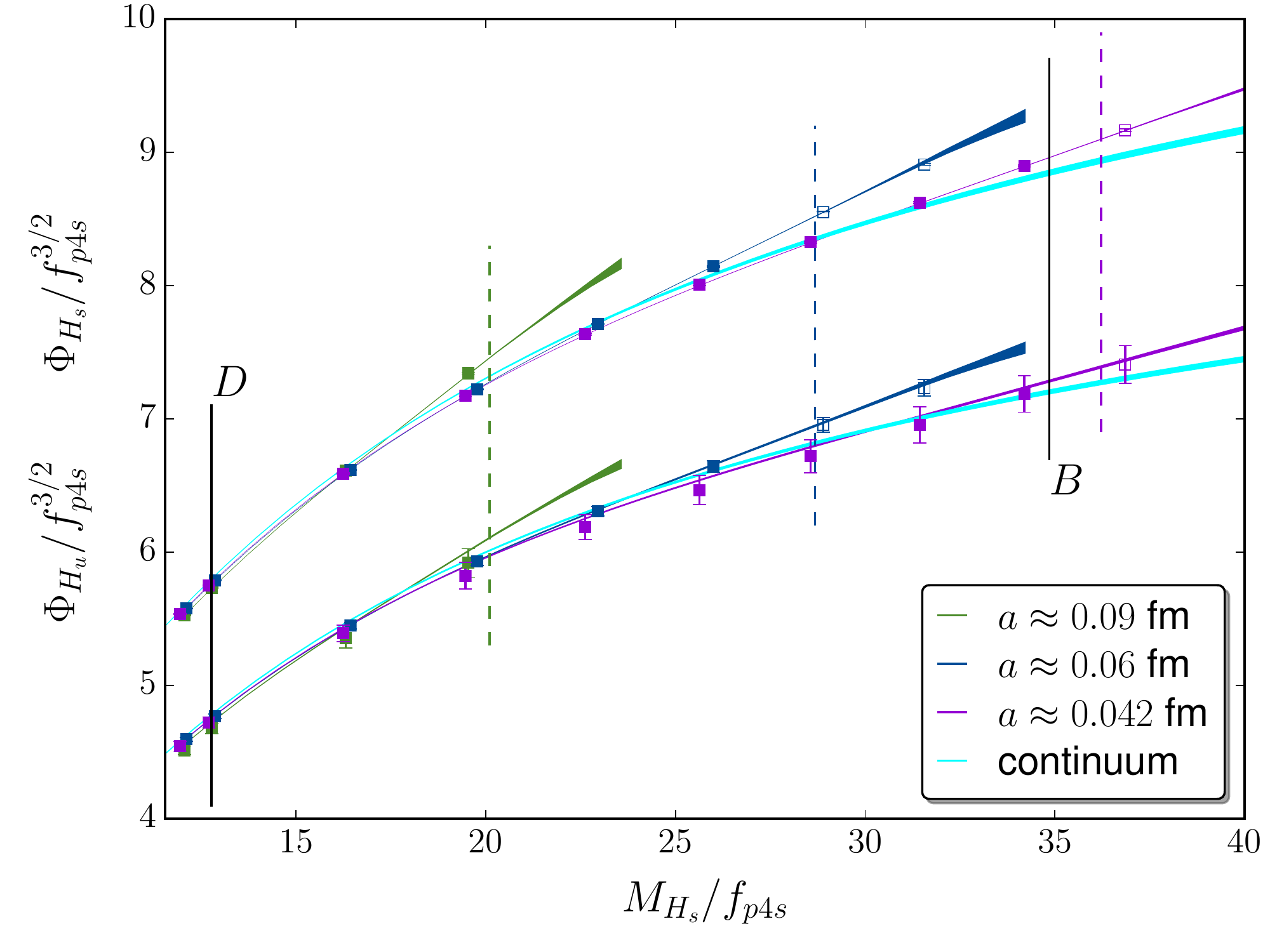}
   \hfill
    \includegraphics[width=2.5in]{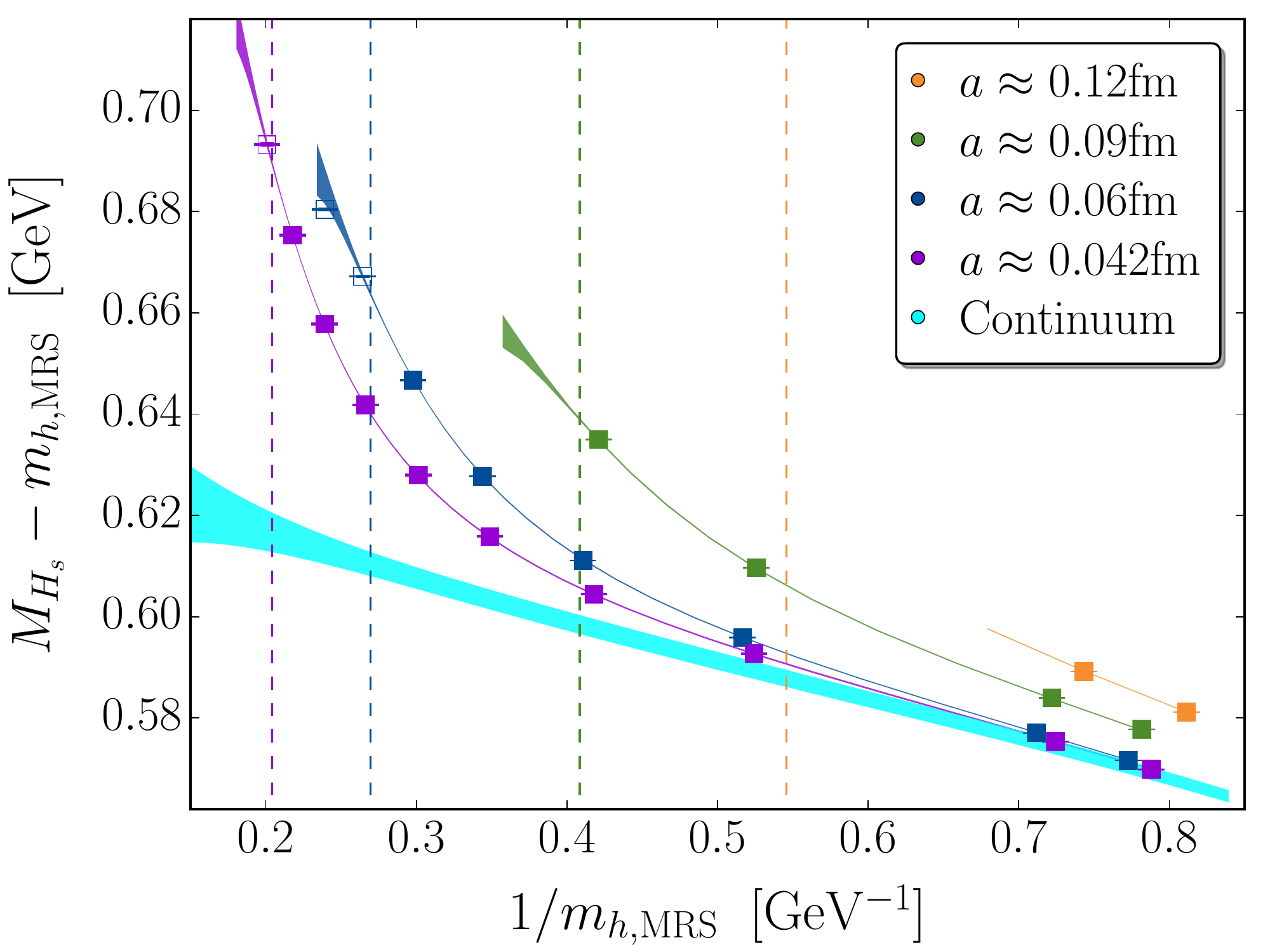}
   \caption{
      Left: snapshot of the base fit and lattice data for the decay
      constant with (lower curves) physical up quark mass and (upper
      curves) physical strange quark mass.  The fit yielded
      $\chi^2/\text{d.o.f} = 466/432$ with $p$ value=0.12.  Right:
      snapshot of the base fit and lattice data for the difference of
      the heavy-strange meson mass and heavy-quark MRS mass {\it vs.}
      the inverse of that quark mass.    The fit yielded 
      $\chi^2/\text{d.o.f} = 320/307$, $p$ value=0.3.
      For both panels the
      dashed lines indicate the $am_h=0.9$ cutoff for the various
      lattice spacings and the cyan bands show the
      extrapolation to zero lattice spacing.
     \label{fig:SampleFits}}
\end{figure}

We set the lattice scale through the well-controlled intermediate
quantity $f_{p4s}$, the decay constant of a fictitious pseudoscalar
meson with both valence masses equal to $m_{p4s} \equiv 0.4m_s$.  Its
physical value is, in turn, determined from $f_\pi$.  This strategy
results in a precise determination of both the lattice spacing {$a$}
and the quark mass {$am_{p4s}$} and in turn $m_s=2.5 m_{p4s}$. The
values of $f_{p4s}$ and quark mass ratio $m_s/m_l$ are determined by
analyzing light-light meson data from the same ensembles.  Various
systematic errors (such as finite volume and electromagnetic effects,
continuum extrapolation \etc) in the estimate of $f_{p4s}$ and tuned
quark masses are incorporated to our estimate of uncertainties

We calculate the decay constant on each ensemble as a function of
lattice spacing and the valence and sea-quark masses.  We fit the
result to a model that eventually permits an
interpolation/extrapolation to the physical point at zero lattice
spacing.  We use a cascade of EFTs to construct the fit functions. We
start from the following schematic form for decay constants of $H_\vt$
mesons
\begin{equation}
   f_{H_\vt}\sqrt{M_{H_\vt}} \equiv \Phi_{H_\vt} = C\, \left(1 + \text{SET} \right)\,\left(1 +    \text{HQET}\right)\,
    \left(1+\text{HMrA\schpt}\right)
    \left(\frac{m'_c}{m_c}\right)^{3/27}\tilde{\Phi}_0 \,.
\end{equation}
The above terms correspond to different effective field theories:
Symanzik effective theory,
\begin{equation}
    \text{SET} = c_1 \alpha_s\, (a\Lambda_{\rm QCD})^2 + \cdots + c_3 \alpha_s\, (am_h)^2 
    + \cdots \,, \nonumber
 \end{equation}
Wilson coefficient,
\begin{equation}
    C =  \Bigl[\alpha_s(M_{H_s})\Bigr]^{-6/25} \Bigl(1 + {\cal O}(\alpha_s) \Bigr) \,, \nonumber
\end{equation}
HQET (and integrating out sea-charm),
\begin{equation}
  \text{HQET} = k_1\frac{\Lambda_{\rm HQET}}{M_{H_s}} 
  + \cdots + k'_1\frac{m_c}{m'_c} \quad \quad
\mbox{and} \quad \quad
  \frac{\Lambda_{\rm QCD}^{(3)}(m'_c)}{\Lambda_{\rm QCD}^{(3)}(m_c)} \approx
  \left(\frac{m'_c}{m_c}\right)^{2/27} \,, \nonumber
\end{equation}
and heavy-meson rooted partially-quenched all-staggered chiral perturbation
theory (HMrA\schpt)\cite{Bernard:2013qwa}
\begin{equation}
  \text{HMrA\schpt} = \mbox{NLO nonanalytic terms} + L_\vt m_\vt
   + L_\text{s} (2m^\prime_l + m^\prime_s) + L_a a^2 \,. \nonumber
\end{equation}
(We use primes to distinguish simulation values from physical values
where ambiguities may occur.) Chiral terms contain effects of taste
splittings, hyperfine and flavor splittings, and finite lattice size.
    
To take into account SET, higher-order \chpt\ effects, and higher order
HQET effects, we include analytic terms.  They are typically
polynomials in dimensionless, ``natural'' expansion parameters.  They
model
\begin{itemize}
\item Light-quark and gluon discretization (SET): $(a\Lambda_{\rm QCD})^2$
      with $\Lambda_{\rm QCD} = 600$ MeV,
\item Heavy-quark discretization effects (SET-HISQ): $(2am_h/\pi)^2$,
\item Light valence and sea quark mass effects (\chpt): $B_0/(4 \pi^2 f_{\pi}^2)\,m_q$,
\item HQET: $\Lambda_{\rm HQET}/M_{H_s}$.
\end{itemize}
The coefficients of the polynomials are fit parameters.  They are
expected to be ${\cal O}(1)$.  Altogether, there are 60 fit parameters
for 492 data points. Figure~\ref{fig:SampleFits} gives a slice of
the fit and data.

\section{Results for decay constants}

We obtain

\begin{align}
    f_{D^0}  &= 211.6  (0.3)_\text{stat} (0.5)_\text{syst} (0.2)_{\fpiPDG} [0.2]_{\rm EM-scheme} \label{eq:fD0}~\MeV,\\ 
    f_{D^+}  &= 212.7  (0.3)_\text{stat} (0.4)_\text{syst} (0.2)_{\fpiPDG} [0.2]_{\rm EM-scheme} \label{eq:fD+}~\MeV,\\ 
    f_{D_s}  &= 249.9  (0.3)_\text{stat} (0.2)_\text{syst} (0.2)_{\fpiPDG} [0.2]_{\rm EM-scheme} \label{eq:fDs}~\MeV,\\ 
    f_{B^+}  &= 189.4  (0.8)_\text{stat} (1.1)_\text{syst} (0.3)_{\fpiPDG} [0.1]_{\rm EM-scheme} \label{eq:fB+}~\MeV,\\ 
    f_{B^0}  &= 190.5  (0.8)_\text{stat} (1.0)_\text{syst} (0.3)_{\fpiPDG} [0.1]_{\rm EM-scheme} \label{eq:fB0}~\MeV,\\ 
    f_{B_s}  &= 230.7  (0.8)_\text{stat} (1.0)_\text{syst} (0.2)_{\fpiPDG} [0.2]_{\rm EM-scheme} \label{eq:fBs}~\MeV.
\end{align}
The systematic error includes uncertainties in the continuum
extrapolation, finite volume correction, the electromagnetic
contribution to meson masses used to fix the quark masses, and the 
adjustment for non-equilibration of topological charge.

These results are obtained in a specific scheme that matches QCD+QED to
pure QCD for the light and heavy meson masses. When using our results
in a setting that does not take into account the subtleties of the EM
scheme, one may wish to also include the estimates of
scheme-dependence given in the last quantities, in brackets.

\begin{figure}
    \includegraphics[width=3in]{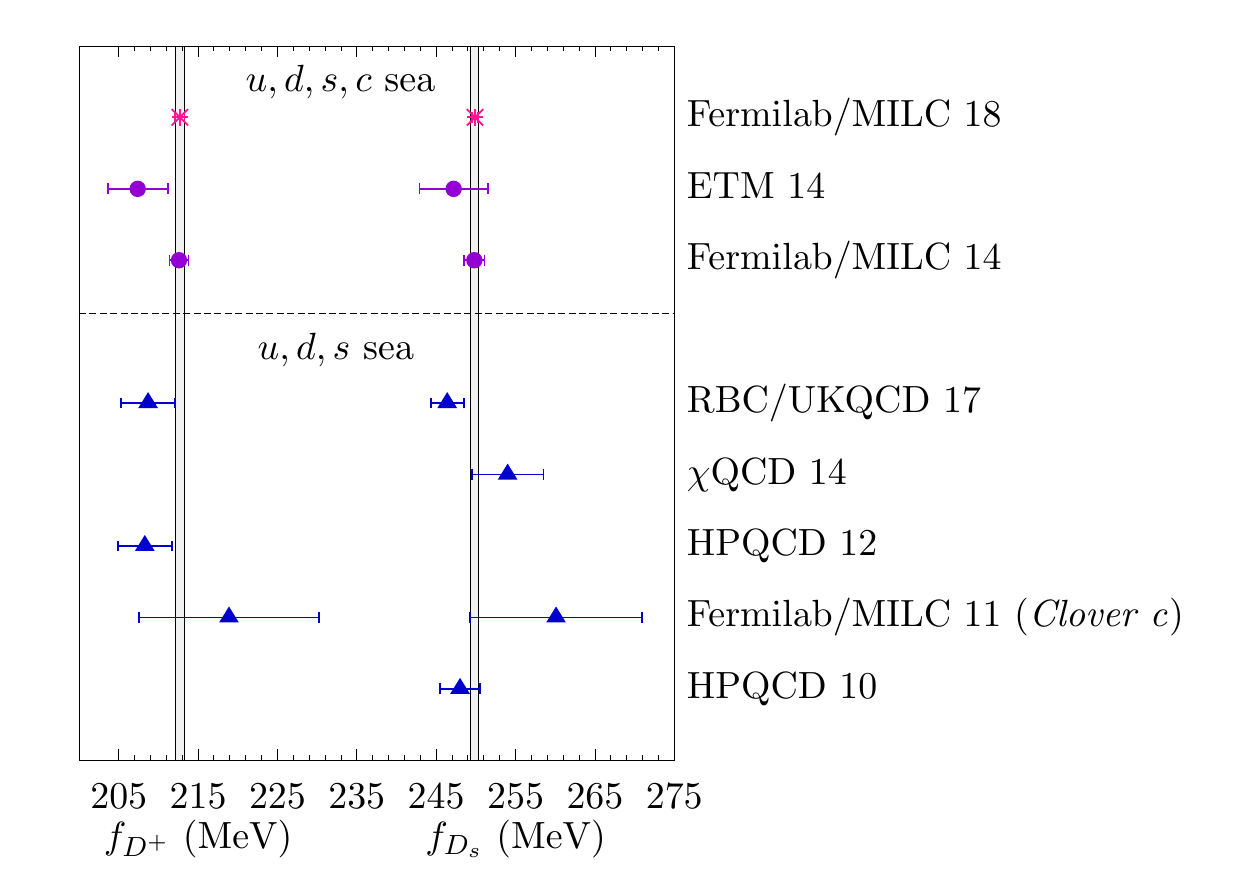}
   \hfill
    \includegraphics[width=3in]{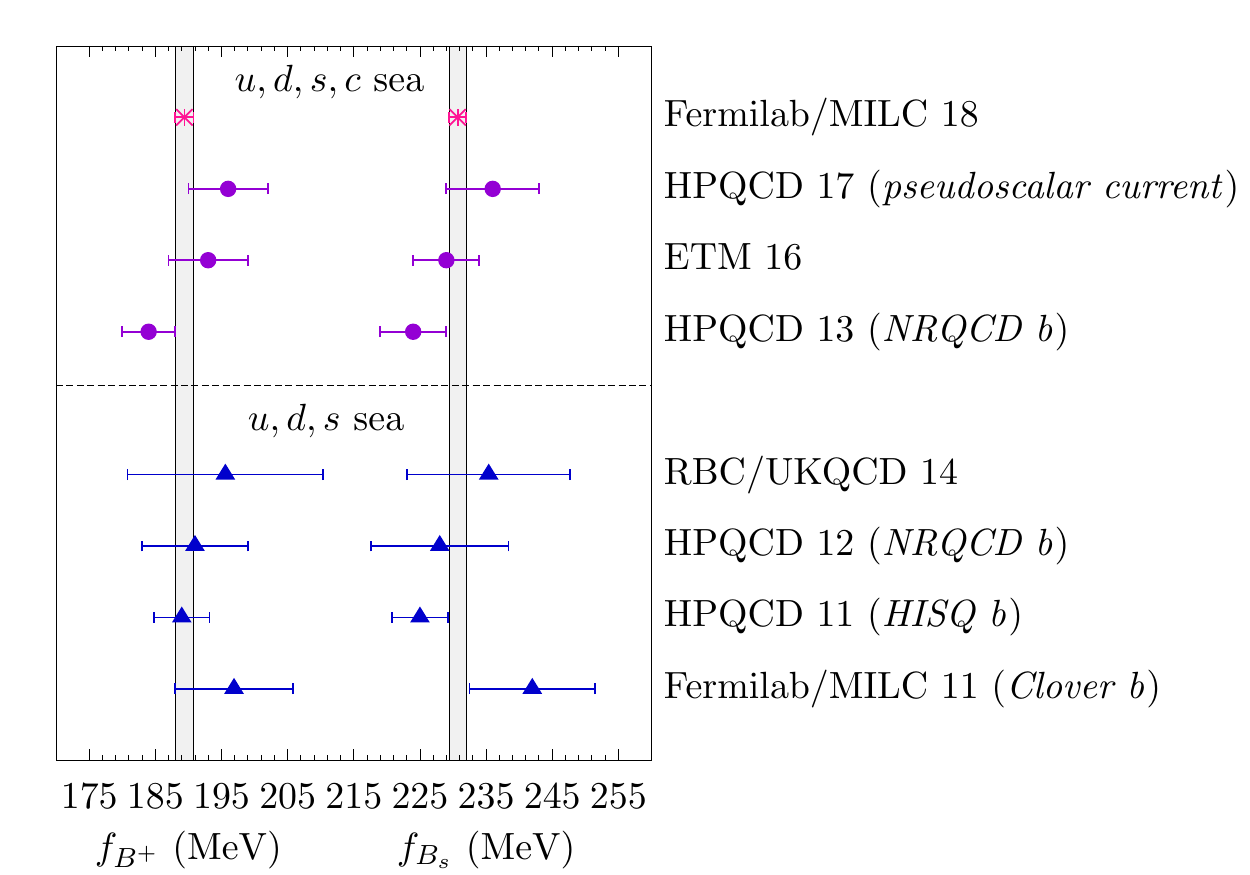}
   \caption{Comparison of our results (magenta) with previous three- and
     four-flavor lattice-QCD calculations.  Labels are defined with
     citations in \cite{Bazavov:2017lyh}.
     %
     %
     The gray bands indicate the total error.
     \label{fig:decaycompare}}
\end{figure}

In Fig.~\ref{fig:decaycompare} we compare some of these results
with those of previous three and four-flavor calculations.

\section{Quark-mass methodology}

The quark masses are determined from the masses of the mesons that
contain them.  For the charm and bottom quarks we use a new method
based on HQET to extract masses of quarks from masses of heavy-light
mesons, starting from a decomposition of the mass:
\begin{equation}
  M_{H} = m_h + \bar{\Lambda} + \frac{\mu_\pi^2 - \mu_G^2(m_h)}{2 m_h}
  + \cO(1/m_h^2) \, ,
\end{equation}
where the parameters are
\begin{itemize}
\item $\bar{\Lambda}$: energy of light quark and gluons inside the system
\item ${\mu_\pi^2}/{2 m_h}$: kinetic energy of the heavy quark inside the system
\item ${\mu_G^2(m_h)}/{2 m_h}$: hyperfine energy due to heavy quark's spin.
  (can be estimated from $B^*$-$B$ splitting: $\mu_G^2(m_b) \approx 0.35\,{\rm GeV}^2$ )
\item $m_h$,  the pole mass of the heavy quark.
    The conventional pole mass is ambiguous because of the renormalon problem.
\end{itemize} 
We use the new {minimal renormalon-subtracted} (MRS) mass introduced
by \cite{Brambilla:2017hcq}.  It removes the leading infrared
renormalon from the pole mass.  It is a gauge- and scale-independent
scheme, and admits a well-behaved perturbative expansion in $\alpha_s$.

In the continuum, the quark mass in the $\MSbar$ scheme is mapped to the
MRS mass using
\begin{equation}
  m_\MRSt = \bar m\left(1 + \sum_{n=0}^\infty [r_n - R_n] \alpha_s^{n+1}(\bar m) + J_\MRSt(\bar m) \right) \,,
  \label{eq:MSbarToMRS}
\end{equation}
where $\bar m \equiv m_\MSbar(m_\MSbar)$ and $J_\MRSt(\bar m)$ is
known \cite{Brambilla:2017hcq}. The small coefficients $[r_n - R_n]$
are the differences between the $\MSbar$ and $\MRSt$ expansion.  We
introduce a ``reference mass'', $m_r = m_{p4s,\MSbar}(\mu)$ and
choose~$\mu = 2\text{GeV}$.

To construct the fit function we begin with the identity,
\begin{equation}
  m_{h,\MRSt} \equiv m_{r,\MSbar}(\mu)\frac{m_{h,\MRSt}}{\bar m_h}\frac{\bar m_h}{m_{h,\MSbar}(\mu)}
  \frac{am_h}{am_r}  \,.
\end{equation}
The first factor is a fit parameter.  The second comes from Eq.~(\ref{eq:MSbarToMRS}) above.
The third comes from $\MSbar$ mass running;
\begin{equation}
  \frac{\bar m_h}{m_{h,\MSbar}(\mu)} = \frac{C(\alpha_\MSbar(\bar m_h))}{C(\alpha_\MSbar(\mu))} \,.
\end{equation}
Finally, the last factor comes from simulation parameters. 

For lattice quantities, the continuum relation above must be modified
to include discretization effects and the light-quark mass dependence.
These terms are obtained through HMrA\schpt\,
\begin{equation}
  M_H = m_{h,\MRSt} + \bar{\Lambda}_\MRSt  + \frac{\mu_\pi^2 - \mu_G^2(m_h)}{2 m_{h,\MRSt}}
  + \text{HMrA\schpt} + \text{higher order HQET} \,.
\end{equation}
As with the decay constants, all terms get corrections for
light-quark- and gluon-discretization effects, based on polynomials in
the natural expansion parameters.  Polynomial coefficients are fit
parameters.  Altogether, there are 67 parameters (6 with external
priors) for 384 data points.

The heavy quark masses are obtained by interpolation to the physical
masses of the heavy-light mesons.  Since electromagnetic effects have been
omitted in the lattice calculation, the physical heavy-light meson masses
must be adjusted before matching to the continuum extrapolation of the fit result.
Light-quark masses and decay constants use light-quark r\schpt\
\cite{Lee:1999zxa,Aubin:2003mg}.

Figure~\ref{fig:SampleFits} gives a slice of the fit and data.
After extrapolating to continuum and matching to measured $M_{D_s}$
and $M_{B_s}$ masses {\em with EM effects removed}, we determine the
strange-quark mass, the charm- and bottom-quark mass ratios, and their masses:
\begin{equation}
  m_{s,\MSbar}(2~\GeV) = 92.47 (39)_\text{stat} (18)_\text{syst} (52)_{\alpha_s} (11)_{\fpiPDG}~\MeV \,,
\end{equation}
\begin{align}
    m_c/m_s  &= 11.783 (11)_\text{stat} (21)_\text{syst} (00)_{\alpha_s} (08)_{\fpiPDG} \,,
    \label{eq:mc_ms}\\
    m_b/m_s  &= 53.94   (6)_\text{stat}  (10)_\text{syst}  (1)_{\alpha_s}  (5)_{\fpiPDG} \,,
    \label{eq:mb_ms}\\
    m_b/m_c  &=  4.578  (5)_\text{stat}  (6)_\text{syst}  (0)_{\alpha_s}  (1)_{\fpiPDG} \,,
    \label{eq:mb_mc}
\end{align}
\begin{align}
    \mbar_c^{(n_f=4)}  &= 1273  (4)_\text{stat} (1)_\text{syst} (10)_{\alpha_s}  (1)_{\fpiPDG}~\MeV \,,
    \label{eq:mc_SI}\\
    \mbar_b^{(n_f=5)}  &= 4195 (12)_\text{stat} (1)_\text{syst}  (8)_{\alpha_s}  (1)_{\fpiPDG}~\MeV \,.
    \label{eq:mb_SI}
\end{align}
where $\mbar_h = m_{h,\MSbar}(m_{h,\MSbar})$. 

The systematic error includes uncertainties in the determination of
scale setting quantities, quark mass tuning, continuum extrapolation,
finite volume, estimating electromagnetic effects.  Shown separately
is the uncertainty in the strong coupling constant, which we take to
be $\alpha_s^{\MSbar}(5\,{\rm GeV}; n_f=4) = 0.2128(25)$ \,
\cite{Chakraborty:2014aca}, and the uncertainty in the value of
$f_\pi$.

\begin{figure}
    \includegraphics[trim={6.5em 0em 1em 1em},clip,width=0.495\textwidth]{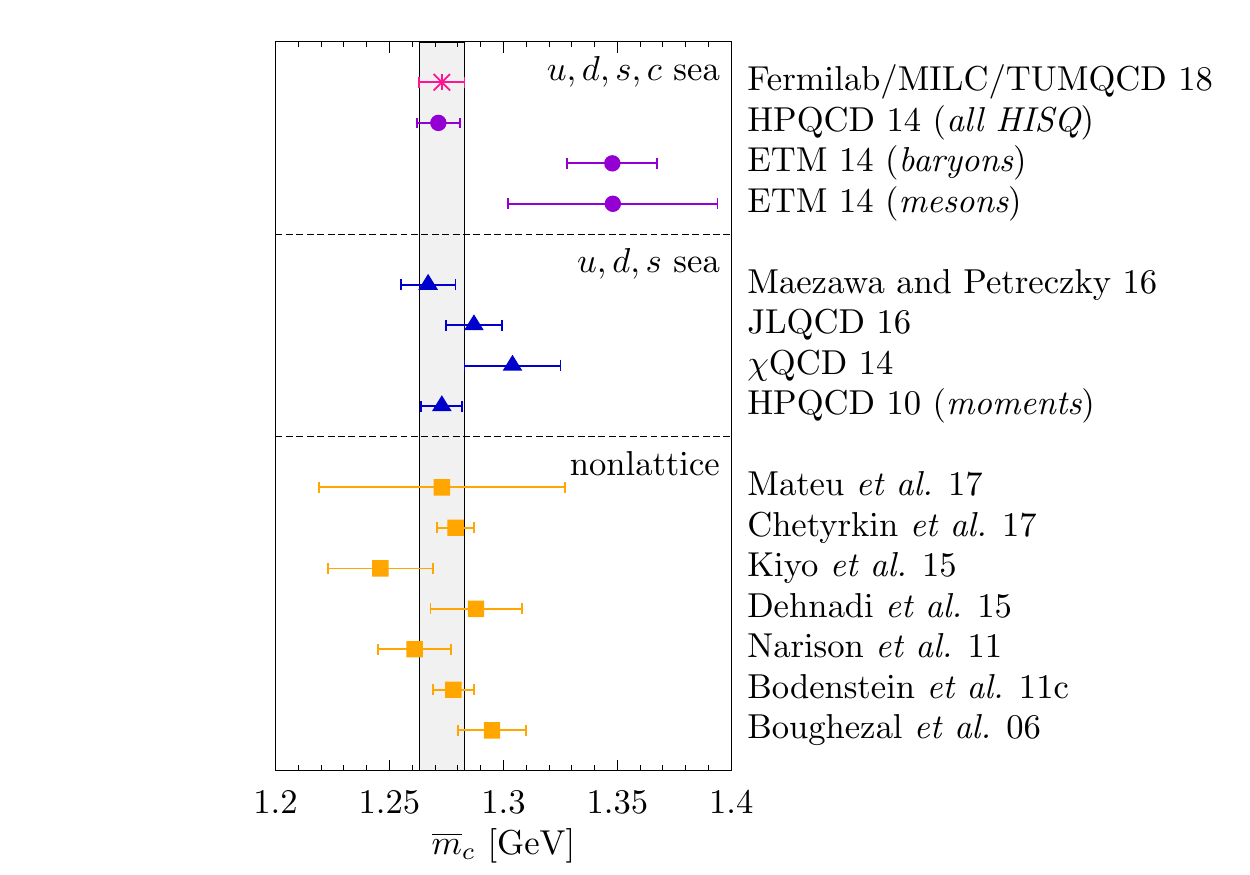} \hfill
    \includegraphics[trim={6.5em 0em 1em 1em},clip,width=0.495\textwidth]{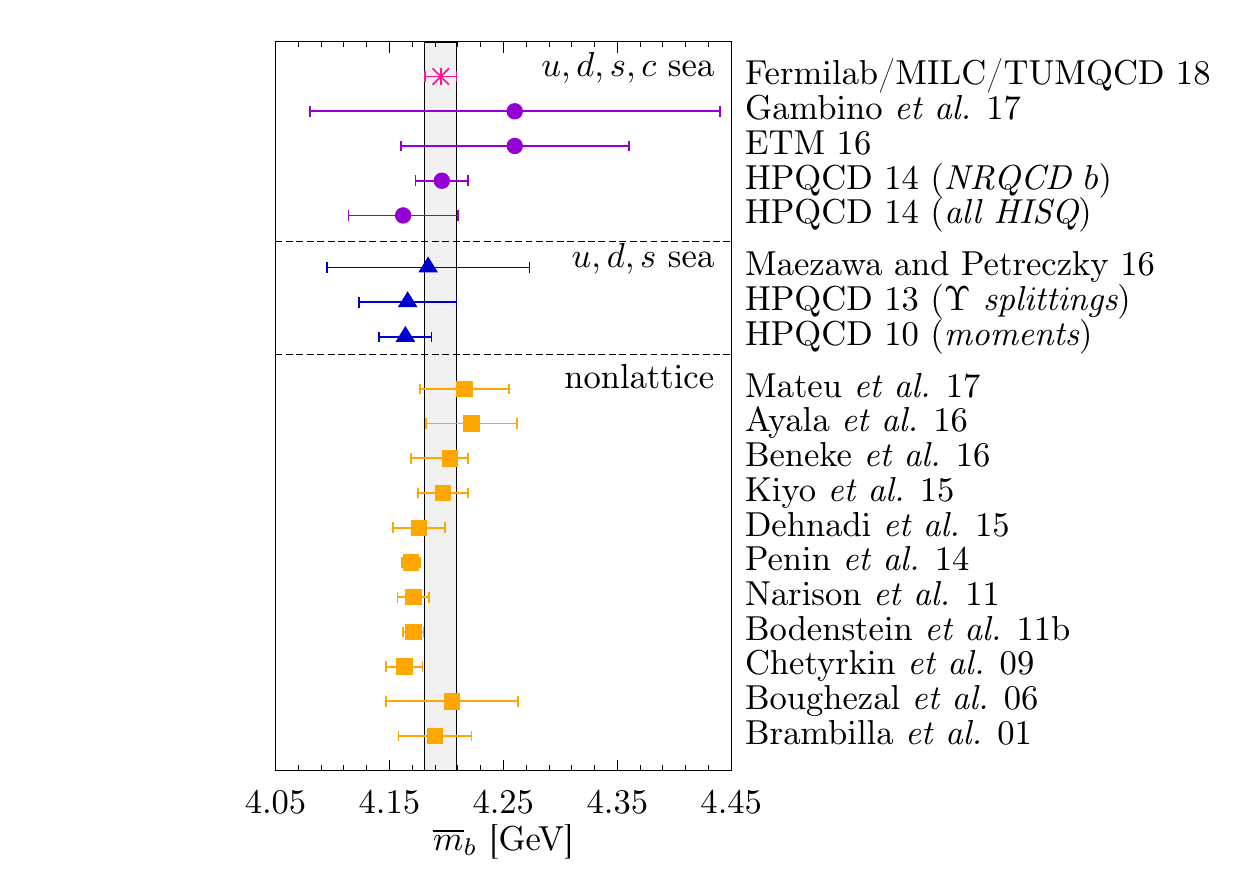}
    \caption{A comparison of our results (in magenta) with those of other
      groups. Labels are defined with citations in \cite{Bazavov:2018omf}.
      The gray bands indicate the total error.
      \label{fig:masscompare}}
\end{figure}

\begin{figure}
    \includegraphics[trim={6em 0em 0em 1em},clip,width=0.495\textwidth]{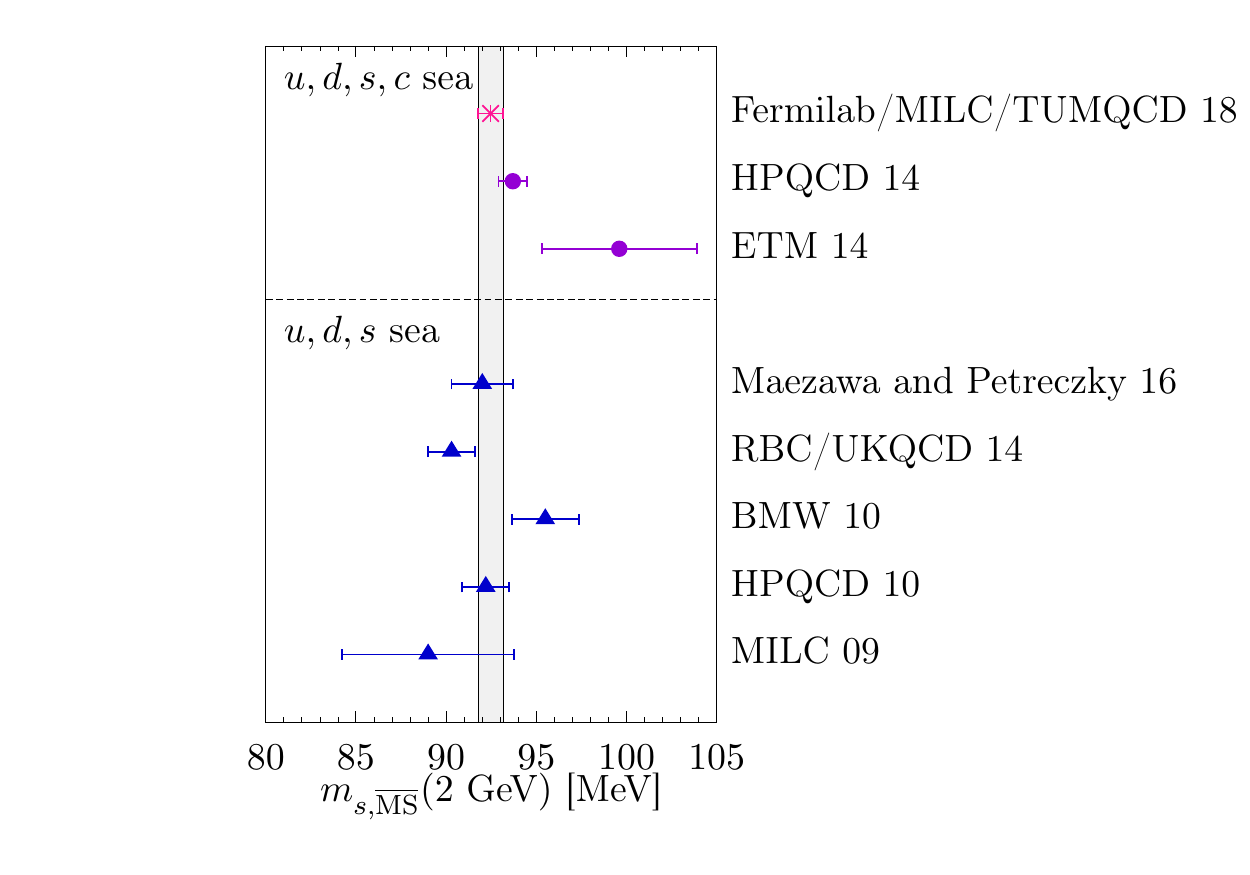} \hfill
    \includegraphics[trim={6em 0em 0em 1em},clip,width=0.495\textwidth]{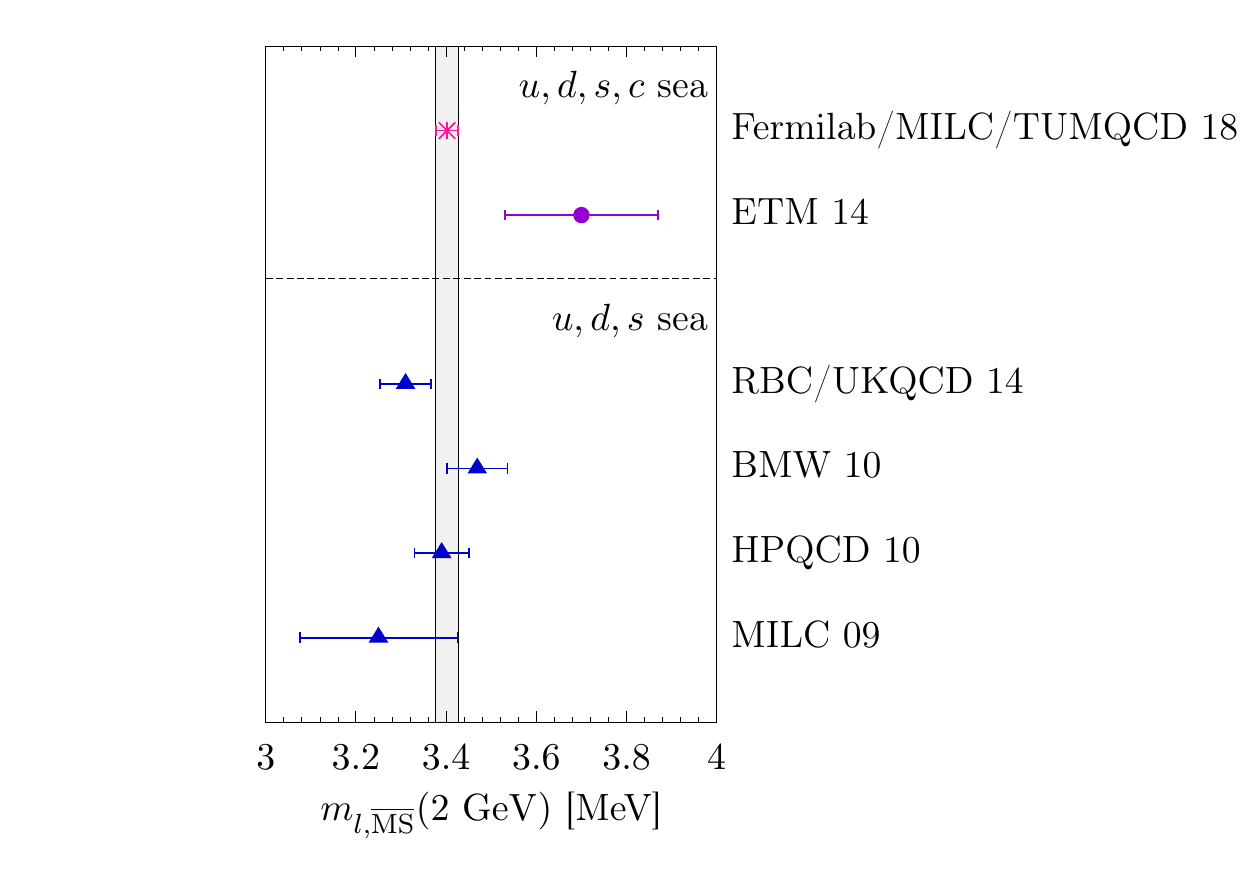}
    \caption{A comparison of our results (in magenta) with those of other
      other lattice QCD calculations. Labels are defined with citations in
      \cite{Bazavov:2018omf}.  The gray bands indicate the total
      error.
      \label{fig:lightmass}}
\end{figure}
Light quark masses are determined from a study of light-light
pseudoscalar meson masses \cite{Basak:2018yzz}:
\begin{align}
    m_{u,\MSbar}(2~\GeV) &= 2.130 (18)_\text{stat} (35)_\text{syst} (12)_{\alpha_s} (03)_{\fpiPDG}~\MeV \,,
    \label{eq:mu_MS} \\
    m_{d,\MSbar}(2~\GeV) &= 4.675 (30)_\text{stat} (39)_\text{syst} (26)_{\alpha_s} (06)_{\fpiPDG}~\MeV \,.
    \label{eq:md_MS}
\end{align}
Our results for the bottom, charm, strange, and the average of the up and down quark masses
are compared with results of other groups in Figs.~\ref{fig:masscompare} and~\ref{fig:lightmass}.

\section{Conclusion}

In this lattice QCD study of the decay constants of heavy-light
mesons, we used a combination of effective field theories in a
correlated, multidimensional fit to lattice data at multiple lattice
spacings.  This approach reduces statistical errors and provides
control of the systematic errors of extrapolation.  We presented
results for decay constants $f_{D^+}$, $f_{D_s}$, $f_{B^+}$, and
$f_{B_s}$.  For quark masses, we developed a method based on
heavy-quark effective theory and the minimal renomalon subtraction
scheme to extract quark masses from masses of the heavy-light mesons.
Again, we used a combination of effective field theories to fit those
heavy-light meson masses.  We presented results for all quark masses
and their ratios.

\Acknowledgements

Computations for this work were carried out with resources provided by
the USQCD Collaboration,
the National Energy Research Scientific Computing Center,
the Argonne Leadership Computing Facility, 
the Blue Waters sustained-petascale computing project,
the National Institute for Computational Science,
the National Center for Atmospheric Research,
the Texas Advanced Computing Center,
and Big Red II+ at Indiana University.
USQCD resources are acquired and operated thanks to funding from the
Office of Science of the U.S. Department of Energy.

This work was supported in part by the U.S.\ Department of Energy under grants
No.~DE-FG02-91ER40628 (C.B., N.B.),
No.~DE-FC02-12ER41879 (C.D.),
No.~DE{-}SC0010120 (S.G.),     
No.~DE-FG02-91ER40661 (S.G.),
No.~DE-FG02-13ER42001 (A.X.K.),
No.~DE{-}SC0015655 (A.X.K.), 
No.~DE{-}SC0010005 (E.T.N.),
No.~DE-FG02-13ER41976 (D.T.);
by the U.S.\ National Science Foundation under grants
PHY14-14614 and PHY17-19626 (C.D.),
PHY14-17805~(J.L.),
and PHY13-16748 and PHY16-20625 (R.S.);
by the MINECO (Spain) under grants FPA2013-47836-C-1-P and FPA2016-78220-C3-3-P (E.G.);
by the Junta de Andaluc\'{\i}a (Spain) under grant No.\ FQM-101 (E.G.);
by the German Excellence Initiative and the European Union Seventh Framework Program under grant agreement No.~291763 as well as 
the European Union's Marie Curie COFUND program (J.K., A.S.K.), and
by the UK Science and Technology Facilities Council (J.K.).

\end{document}